\documentclass[12pt,pra,reprint,superscriptaddress,amsmath,amsfonts,amssymb]{revtex4-1}

\usepackage[pretty]{revquantum}
\usepackage[english]{babel}
\usepackage{graphicx}
\usepackage[caption=false]{subfig}
\usepackage{mathrsfs}
\usepackage{amsmath}
\usepackage{amsthm}
\usepackage{thmtools, thm-restate}
\usepackage{bm}
\usepackage{xcolor}
\usepackage{hyperref}
\usepackage{hypernat}
\usepackage[utf8]{inputenc}
\usepackage[T1]{fontenc}
\usepackage{cleveref}
\usepackage{soul}

\declaretheorem[parent=section,name=Theorem]{thm}

\declaretheorem[style=definition,sibling=thm]{definition}

\DeclareGraphicsExtensions{.pdf,.png,.jpg}

\begin{document}
	

	
	\title{Bell Non-Locality in Many Body Quantum Systems with Exponential Decay of Correlations}

	\author{Carlos H. S. Vieira}
\affiliation{Departamento de Matem\'{a}tica Aplicada, Instituto de Matem\'{a}tica, Estat\'{i}stica e Computa\c{c}\~{a}o Cient\i fica, Universidade Estadual de Campinas, 13083-859, Campinas, S\~{a}o Paulo, Brazil}
\email[Corresponding author:]{chumbertosvieira@gmail.com}
\author{Cristhiano Duarte}
\affiliation{Schmid College of Science and Technology, Chapman University, One University Dr., Orange, CA 92866, USA}
\affiliation{Institute for Quantum Studies, Chapman University, One University Dr., Orange, CA 92866, USA}
\author{Raphael C. Drumond}
\affiliation{Departamento de Matem\'{a}tica, Instituto de Ci\^{e}ncias Exatas, Universidade Federal de Minas Gerais, 30123-970, Belo Horizonte, Minas Gerais, Brazil}	
\author{Marcelo Terra Cunha}
\affiliation{Departamento de Matem\'{a}tica Aplicada, Instituto de Matem\'{a}tica, Estat\'{i}stica e Computa\c{c}\~{a}o Cient\i fica, Universidade Estadual de Campinas, 13083-859, Campinas, S\~{a}o Paulo, Brazil}

	\date{\today}

	\begin{abstract}
		Using Bell-inequalities as a tool to explore non-classical physical behaviours, in this paper we analyse what one can expect to find in many-body quantum physics. Concretely, framing the usual correlation scenarios as a concrete spin-lattice, we want know whether or not it is possible to violate a Bell-inequality restricted to this scenario. Using clustering theorems, we are able to show that a large family of quantum many-body systems behave almost locally, violating Bell-inequalities (if so) only by a non-significant amount. We also provide examples, explain some of our assumptions via counter-examples and present all the proofs for our theorems. We hope the paper is self-contained.
	\end{abstract}
	
	\maketitle
	
	
	\section{Introduction}

	Quantum physics features correlations showing no parallel with classical physics. 
	Bell non-locality and contextuality being the most prominent examples. 
	The former can be understood as a phenomenon in which the statistics obtained from local measurements acting on distant parts of a quantum system cannot be replicated by any model of (local) classical variables \cite{bell1964einstein}. 
	In other words, the statistics shown by this type of local experiments cannot be reproduced from deterministic strategies, even if aided by shared randomness \cite{Fine82}. 
	The fact that local deterministic strategies fail to frame scenarios exhibiting non-local data is usually detected through violations of so called Bell inequalities \cite{bell1964einstein,CHSH,Bell-Nonlocality}:  linear combinations of expected values of correlations from local measurements with a bound calculated under the assumption of Bell locality. 
	A violation of such inequalities witnesses the presence of Bell non-locality in the system (for a review see \cite{Bell-Nonlocality}).
	
    Ultimately, non-locality is only manifest when considering a scenario involving multiple physical systems, be them  black-boxes in the device independent scenario or actual quantum systems. 
    In particular, non-locality in many-body quantum systems has been extensively explored \cite{batle_nonlocality_2010, batle_nonlocality_2011, campbell_multipartite_2010, wang_entanglement_2017, oudot_bipartite_2019, sun_multipartite_2019, sun_multipartite_2019-1, Batle-Casas-QPT, Altinas-Ferdi-QPT, Justino-Thiago-QPT, Deng-Wu-QPT, thiago-nonviolation, SUN201430, GETELINA20182799}, see \cite{De_Chiara_2018} for a review. For instance, it has been discussed in the literature how to use non-locality measurements as an indicator of quantum phase transitions (QPT's) in several many-body systems models \cite{Batle-Casas-QPT, Altinas-Ferdi-QPT, Justino-Thiago-QPT, Deng-Wu-QPT}. 
    In all of these works, Bell correlations between spin pairs, measured through CHSH inequality \cite{CHSH}, were used to characterize QPT's. 
    Surprisingly, it was observed that such inequality was not violated in any of these models \cite{Batle-Casas-QPT, Altinas-Ferdi-QPT, Justino-Thiago-QPT, Deng-Wu-QPT}.

     As a matter of fact,  considering only the overlap between many-body quantum systems and the use of CHSH-violation as a marker for quantumness, it is remarkable how rich non-locality is. 
     On one hand, in Ref.~\cite{thiago-nonviolation} the authors showed for translationally-invariant lattices, pairs of spins do not exhibit any violation of CHSH inequality, even though the global state may be highly-entangled. 
     On the other hand, it is known that for simple lattices with no translational symmetry it is, indeed, possible to get CHSH violations for some pairs of sites \cite{SUN201430, GETELINA20182799}.
    
     Detection of multipartite non-locality is another example of the exchange between many-body physics and foundations of quantum mechanics. Although it is known that it is mathematically hard to characterize non-local effects in more complex Bell scenarios  \cite{Babai1991}, recent work has shown that it is possible to detect multipartite non-locality by simpler Bell inequalities, involving only two-body correlators \cite{tura2014detecting, TURA2015370, Tura_2014, PhysRevLett.119.230402, PhysRevLett.118.230401, piga_bell_2019}. 
     In particular, in Ref.~\cite{tura2014detecting} is demonstrated that physically relevant states, such as the ground state of some spin models in many-body systems, exhibit non-locality for these types of Bell inequality. 
     In Ref.~\cite{PhysRevX.7.021005}, it was remarked that some observables from many-body systems, like energy, can be used as a witness to non-locality. From these tools it was possible to witness non-locality in a Bose-Einstein Condensate of 480 atoms \cite{schmied2016bell} and in a thermal ensemble of $5\times10^{5}$ atoms \cite{PhysRevLett.118.140401}.
    
    This work is placed exactly at this intersection between foundations of physics and  many-body quantum mechanics. 
    As a matter of fact, ihere we investigate general non-local features in spin lattices. 
    More precisely,  we will show two situations in which regions of the system can not show expressive non-locality when measurements are made in sufficiently distant regions of the lattice:  ground states of gapped Hamiltonians and  thermal equilibrium states of these latices for high temperature. 
    We also analyze how violations of Bell inequalities can arise from the interactions of the spins in a lattice, when the initial state is product. 
  
    The paper is organized as follows: In Section \ref{Sec.Results} we present our main results followed by a short discussion. 
    In Section \ref{Sec.Preliminaries}  we give a short review on the necessary aspects of nonlocality and the clustering theorems for many-body Hamiltonians. 
    In Section \ref{Sec.Proofs} we give the proofs of the results enunciated in Section \ref{Sec.Results}, before 
 conclusions are shown together with a discussion of future lines of research, in Section \ref{Sec.Conclusion}.
	
	\section{Results}\label{Sec.Results}

	This section contains the main results of our work. Every definition, lemma and theorem is followed or come right after a short motivation or justification. This way we feel this section can stand by itself.
	
	 However, we are bridging between two quite well-established fields, so that we are building our findings upon some common knowledge and jargon coming from many-body quantum systems and foundations of quantum mechanics. If the reader is not comfortable with the presentation, we refer them to Section \ref{Sec.Preliminaries} where we present the basics necessary for a better hold of our results.  
	\subsection{Main Results}
	The simplest Bell scenario is one in which two causally-separated agents, Alice and Bob, have available two dichotomics measurements each. Alice has access to $A_{0}$, $A_{1}$ and Bob has access to $B_{0}$, $B_{1}$ \cite{Bell-Nonlocality}. Up to relabeling, the only non-trivial Bell inequality for this scenario is the CHSH inequality \cite{CHSH}:
	\begin{equation}\label{CHSH}
	\left<A_{0}B_{0}\right>+\left<A_{0}B_{1}\right>+\left<A_{1}B_{0}\right>-\left<A_{1}B_{1}\right>\le 2.
	\end{equation}

	In this Bell scenario, every system exhibiting an aggregated statistics verifying the inequality in \eqref{CHSH} is called \emph{local} and the correlations presented by it can be explained by a local theory \cite{Bell-Nonlocality}. Non-local quantum features are already manifest even at this simple scenario, as we know this inequality can be violated by a particular choice of measurements and states,  with the maximum violation reaching $2\sqrt{2}$ \cite{cirel1980quantum}.
	
	We can realize this Bell experiment via a quantum spin system. Let $\Omega$ be a lattice representing the location of a finite set of spins, and let $\mathcal{H}_{\Omega}$ be the Hilbert space associated with that lattice. Additionally, consider that the spins interact with each other, this interaction given by a Hamiltonian operator $H$ acting on $\mathcal{H}_{\Omega}$. We will assume that the interactions are short-ranged, that is, the range of the interactions is small compared to the size of the lattice. In this experiment, Alice has her action restricted to a region $X\subset\Omega$ of the system while Bob has his action restricted to a region $Y\subset\Omega$, as illustrated in \textbf{Figure \ref{lattice}}. Denote by $r$ the distance between $X$ and $Y$, and by $|Z|$ the number of sites in a region $Z\subset\Omega$. Alice's measurements are operators acting on the lattice with support in the region $X$ while Bob's  measurements are also operators acting on the lattice but supported in $Y$. Finally, assume also that the norm of these operators are upper-bounded by 1.
	
	In this setting, the expected values of these measurements are given by $\left<A_{i}B_{j}\right>_{\rho}=Tr(\rho A_{i}B_{j})$ where $\rho$ is the state of the whole spin system. Thus, denoting $\textbf{A}=\{A_0,A_1\}$ and $\textbf{B}=\{B_0,B_1\}$ we can define the following quantities.
	\begin{align}\label{CHSH in lattice}
	\mathcal{B}_{CHSH}^{X,Y}(\rho,\textbf{A},\textbf{B})&:=\left<A_{0}B_{0}\right>_{\rho}+\left<A_{0}B_{1}\right>_{\rho}    \nonumber\\
	&+\left<A_{1}B_{0}\right>_{\rho}-\left<A_{1}B_{1}\right>_{\rho};
	\end{align}
	\begin{equation}\label{sup CHSH in lattice}
	\mathcal{B}_{CHSH}^{X,Y}(\rho)=\sup_{\textbf{A},\textbf{B}}\mathcal{B}_{CHSH}^{X,Y}(\rho,\textbf{A},\textbf{B}),
	\end{equation}
	where we are optimising over all operators $A_{i},B_{i}$  acting on $X$ and $Y$ with $\Vert A_i \Vert, \Vert B_i \Vert  \leq 1$. Therefore, if $\mathcal{B}_{CHSH}^{X,Y}(\rho)\le2$ the state $\rho$ is local for this Bell experiment.  
	
	Our  goal is to use clustering theorems to recover almost local behaviours for many-body quantum systems. We want to guarantee that when the two parts $X,Y$ are far away from each other, regardless of the rest of the system, possible violations of CHSH are vanishingly small. For doing so, we define the following class of states.
    \begin{definition}\label{Def.EpsilonLocal}
    Given two disjoint  regions $X,Y\subset\Omega$ and a real number $\epsilon>0$, a quantum state $\rho$ acting on $\mathcal{H}_{\Omega}$ is $\epsilon$\emph{-local} with respect to CHSH and with relation to these two regions if 
    \begin{equation}\label{Eq.DefEpsilonLocal}
      \mathcal{B}_{CHSH}^{X,Y}(\rho)\le2+\epsilon.
    \end{equation}   
    \end{definition}
    It is important to note that the notion of $\epsilon$-locality defined above is linked to the $X,Y$ regions. What we are going to show, though, is that there are important classes of $\epsilon$-local states, with $\epsilon \ll 1$ regardless of regions, as long as they are sufficiently separated from each other. Actually, that is a quite natural assumption in Bell experiments, as assuming the agents are far from each other ensures that there is no direct causal influence on the correlations.
    
    The above discussion motivates the definition of a state states with exponential clustering of correlation \cite{communications,Nachtergaele2006-2,kliesch2014locality}: 
    \begin{figure}
		\begin{center}
			\includegraphics[scale=0.4]{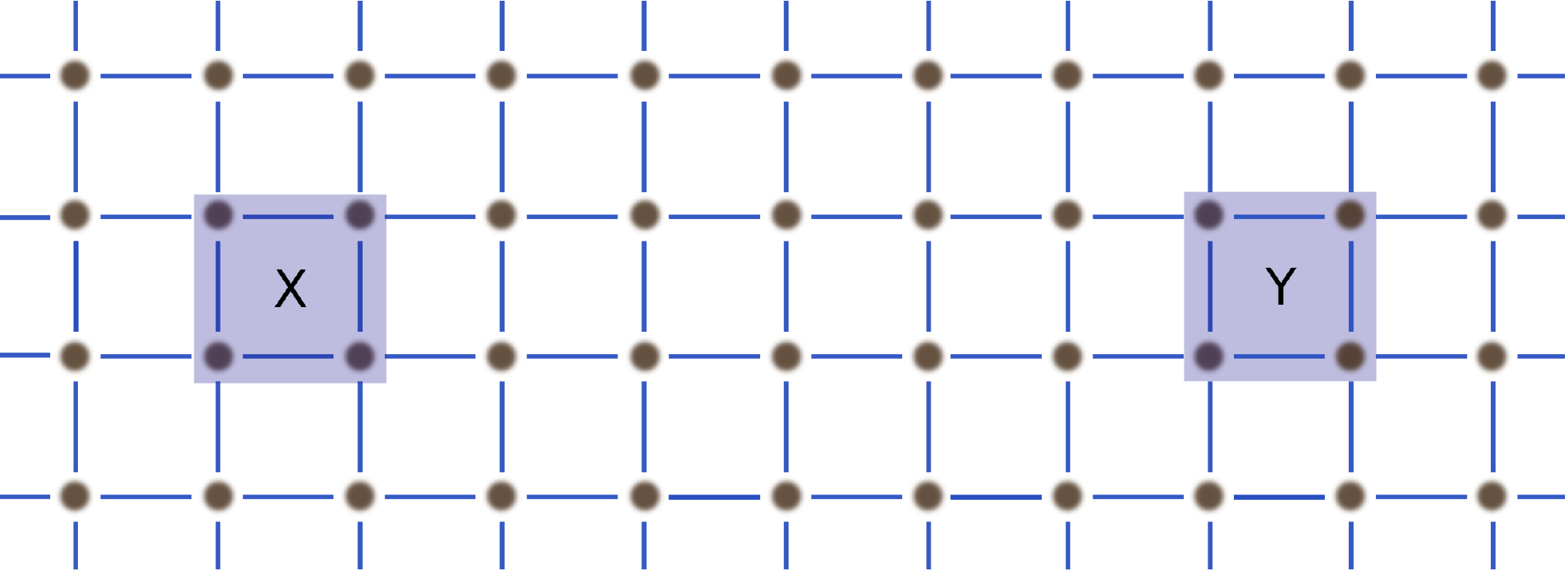}
			\caption{Bell experiment in a spin system}
			\label{lattice}
		\end{center}
	\end{figure}
    \begin{definition}\label{state with ecc}
    	A quantum state $\rho$ acting on $\mathcal{H}_{\Omega}$ shows \textit{exponential clustering of correlations} if there are two positive constants $C,\lambda$,  so that for any two disjoint regions $X,Y$ and any pair of operators $A,B$ supported at $X,Y$ respectively, we have 
    	\begin{equation}\label{eq state with ecc}
               \left|\left<AB\right>_{\rho}-\left<A\right>_{\rho}\left<B\right>_{\rho}\right|\le \|A\| \|B\||X| Ce^{-\lambda r}.
    	\end{equation}
    \end{definition}
    
    \textbf{Remark:} For sake of simplicity and to improve the readability, we are always assuming that $|X| \leq |Y|$. The general case is obtained by changing from $|X|$ to min$\{|X|,|Y|\}$.  As we are more interested in the distance between the the subsets, we will stick to our assumption without any loss of generality.
    
    Two important classes of states that has \textit{exponential clustering of correlations} are the ground state of a gapped Hamiltonian (see   \textbf{Theorem \ref{dec groundstate}} below); and the thermal quantum states at inverse temperature less than a fixed $\beta^{\ast}$  (see \textbf{Theorem \ref{dec thermal states}} below). In fact, theorems of type \ref{dec groundstate}, \ref{dec thermal states} are ussually called by Clustering Theorems \cite{fredenhagen_remark_1985}.

    A state showing \textit{exponential clustering of correlations} has small correlations between distant parts. It is therefore expected that the non-local correlations will also be small, the following lemma assure us of this.
	\begin{lemma}\label{lemma1}
    Given two disjoint regions $X,Y\subset\Omega$ and $\rho$ a quantum state with \textit{exponential clustering of correlations}, then $\rho$ is $\epsilon$-local for CHSH with respect this two regions, where $\epsilon=4|X|Ce^{-\lambda r}$.		
 	\end{lemma}

    Recall that the constants $C,\lambda$ do not depend on the regions. Therefore, by distancing Bob from Alice, so that $r$ becomes increasingly larger, $\epsilon$ will be as close to zero as you want.
         
    As mentioned earlier, \textbf{Theorem \ref{dec groundstate}} ensures \textit{exponential clustering of correlations} for the ground state of a gapped Hamiltonian. So, from \textbf{Theorem \ref{dec groundstate}} and \textbf{Lemma \ref{lemma1}} we get our first main result. 

	\begin{theorem}\label{CHSH for groundstate}
		If $\rho$ is the ground state of a gapped Hamiltonian of the lattice, then there is $C,\lambda>0$ such that given $X,Y\subset\Omega$ disjoint regions we have that $\rho$ is $\epsilon$-local state for CHSH with respect these two regions, where $\epsilon=4|X|Ce^{-\lambda r}$.
	\end{theorem}	
    For the same reasons already presented, we will have a small $\epsilon$ if the distance between the parts is large, as expected in a Bell experiment.

	We also can use \textbf{Theorem \ref{dec thermal states}} and \textbf{Lemma \ref{lemma1}} to show a similar property for thermal states. However, a thermal state has additional properties that allow us to show a stronger result.
	\begin{theorem}\label{CHSH for thermal states}
		Let $\rho(\beta)$ be a thermal state acting on the lattice with a inverse temperature $\beta$ less than a fixed $\beta^{\ast}$, and let $\textbf{A}$ be a set of operators acting on $X\subset\Omega$. There is $r^{\ast}>0$ such that given $Y\subset\Omega$ with $r\ge r^{\ast}$ we have $\mathcal{B}_{CHSH}^{X,Y}(\rho(\beta),\textbf{A},\textbf{B})\le2$ for every set of operators $\textbf{B}$ acting on $Y$.
	\end{theorem}   
    Broadly speaking, \textbf{Theorem \ref{CHSH for thermal states}} is saying that  for every  choice of measurements for Alice, if Bob is far enough, we can not see non-locality in the experiment. 
	
    There is another theorem from many body quantum systems that implies an exponential decay of correlations with the distance (see \textbf{Theorem \ref{prop_cor}} bellow). This theorem bounds the propagation of correlations in the lattice is when we start from a product state. Applying \textbf{Theorem \ref{prop_cor}} and using similar ideas, as in the proof of \textbf{Lemma \ref{lemma1},} we can enunciate the following result.
	\begin{theorem}\label{CHSH for product states}
 	Suppose the initial state of the system is a product state, i.e, $\rho(0)=\otimes_{x\in\Omega}\rho_{x}$. Then, there is $C,v,\lambda>0$ such that given two disjoint regions $X,Y\subset\Omega$ then $\rho(t)$ is $\epsilon$-local for CHSH with respect these two regions, where $\epsilon=4|X||Y|C(e^{\lambda vt}-1) e^{-\lambda r}$.
	\end{theorem}
     The constant $v$ is called the Lieb-Robinson velocity and it represents the maximum effective velocity of propagation of the information across the lattice \cite{LR}. Therefore, we conclude that we will have an effective local behavior for a time of the order $\frac{r}{v}$.
	
    So far, we have used CHSH as a tool for non-locality detection. However, some of the previous results can be promptly generalized to more complex scenarios with richer Bell inequalities. So, let us consider a scenario where $N$ spatially separated agents share a quantum state. Each party $i$ chooses one out of the $m$ possible $M_{i}$ dichotomic measurements, and performs it on his part of the shared quantum state. The reason for restricting it to dichotomic measurements comes from the fact that in this case a Bell inequality can be written through correlators \cite{Bell-Nonlocality}.
    
    A Bell inequality for this scenario involves the sum of correlators between many parts at the same time. However, as discussed in the introduction, there is an interest in Bell inequalities with correlators of at most two bodies. These inequalities are simpler, and from them it will be possible to better visualize our results. We will start from these inequalities and at the end of this section we will return to the general case.
    
     A general Bell inequality involving correlators of one and two bodies can be written as:

	\begin{align}
	&\sum_{i,k=1}^{N,M_{i}}\alpha_{k}^{(i)}\left<E_{k}^{(i)}\right> + \sum_{i\neq j}^{N}\sum_{k,l=1}^{M_{i},M_{j}}\beta_{kl}^{(ij)}\left<E_{k}^{(i)}E_{l}^{(j)}\right>\le \Delta_{C},
	\end{align}
    where $E_{k}^{(i)}$ is the $k-th$ measurement of agent $i$ and  $\alpha_{k}^{(i)}$, $\beta_{kl}^{(ij)}$, $\Delta_{C}$ are real constants, with  $\Delta_{C}$ the called \emph{local bound}. Again, every state whose aggregated statistics respects this inequality is called \emph{local}.
    
     This family of Bell inequalities is already capable of signaling out non-locality for physically relevant states \cite{tura2014detecting}. It is a fact that in a bipartite scenario where all measurements are dichotomic, all inequalities can be written in this way. On the other hand, in a multipartite scenario, this class of inequalities is important due to the ease of implementation in many-body systems models \cite{schmied2016bell,PhysRevLett.118.140401}.
	\begin{figure} 
		\begin{center}
			\includegraphics[scale=0.4]{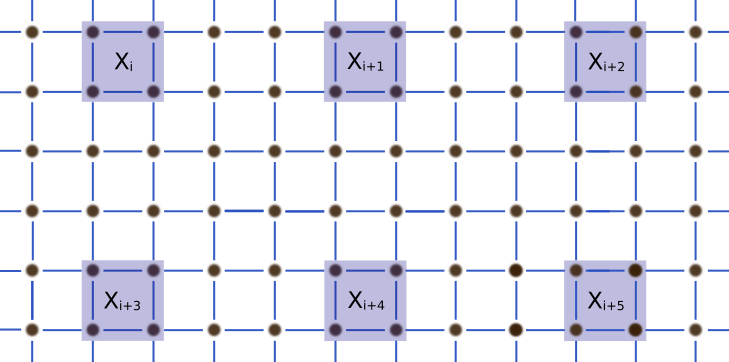}
			\caption{Multipartite Bell experiment in a spin system}
			\label{lattice2}
		\end{center}
	\end{figure}

	Again, we can perform this Bell experiment on a quantum spin system. Now, each agent $i$ has their action restricted to a region $X_{i}$ of the system, as illustrated in \textbf{Figure} \ref{lattice2}. We will indicate by $r_{ij}$ the distance between the regions $X_{i}$ and $X_{j}$. As before, the measurements from the agent $i$ are operators acting on the lattice with support in the region $X_{i}$ and with norm less than or equal to $1$. Let us denote by $\textbf{E}^{(i)}$ the set of measurements operators from agent $i$, that is $\{E_{1}^{(i)},\cdots,E_{M_{i}}^{(i)}\}$. Similarly to \eqref{CHSH in lattice}, \eqref{sup CHSH in lattice}, we define

	\begin{align}\label{Bell bound}
	&\mathcal{B}_{2Body}^{X_{1},...,X_{N}}(\rho,\textbf{E}^{(1)},\cdots,\textbf{E}^{(N)})=\sum_{i,k=1}^{N,M_{i}}\alpha_{k}^{(i)}\left<E_{k}^{(i)}\right>_{\rho}\nonumber\\
	&+\sum_{i\neq j}^{N}\sum_{k,l=1}^{M_{i},M_{j}}\beta_{kl}^{(ij)}\left<E_{k}^{(i)}E_{l}^{(j)}\right>_{\rho};
	\end{align}
	\begin{equation}
	\mathcal{B}^{X_{1},...,X_{N}}_{2Body}(\rho)=\sup_{\textbf{E}^{(1)},\cdots,\textbf{E}^{(N)}}\mathcal{B}^{X_{1},...,X_{N}}_{2Body}(\rho,\textbf{E}^{(1)},\cdots,\textbf{E}^{(N)}).
	\end{equation}
	If $\mathcal{B}^{X_{1},...,X_{N}}_{2Body}(\rho)>\Delta_{C}$ then the state $\rho$ shows non-locality in this configuration of Bell's experiment. The generalization for \textbf{Lemma \ref{lemma1}} is the following.
	\begin{lemma}\label{lemma2}
		Let $\rho$ be a quantum state acting on $\mathcal{H}_{\Omega}$ showing \textit{exponential clustering of correlations}. Then, there exist $C,\lambda>0$ such that for every $X_{1},\cdots,X_{N}\subset \Omega$ disjoint regions we have $$\mathcal{B}^{X_{1},...,X_{N}}_{2Body}(\rho)\le\Delta_{C}+C\sum_{i\neq j}^{N}\sum_{k,l=1}^{M_{i},M_{j}}\min\{|X_{i}|,|X_{j}|\}|\beta_{kl}^{(ij)}|e^{-\lambda r_{ij}}.$$
	\end{lemma}

    The conclusion for \textbf{Lemma \ref{lemma2}} is the same as for \textbf{Lemma \ref{lemma1}}. Again, the constants involved are independent of the regions. Therefore, if all regions are sufficiently distant from each other, we will not be able to see any significant violation in any of these Bell inequalities.

    As we mentioned before, states obey Cluster Theorems also showing exponential clustering of correlation. Then, using \textbf{Lemma \ref{lemma2}} together with the \textbf{Theorem \ref{dec groundstate}}, the following generalization from \textbf{Theorem \ref{CHSH for groundstate}} is obtained. 
	\begin{theorem}\label{twobody bell inequality groudstate}
		If $\rho$ is the ground state of a gapped Hamiltonian of the lattice, then is $C,\lambda>0$ such that for every $X_{1},\cdots,X_{N}\subset\Omega$ disjoint regions we have $$\mathcal{B}^{X_{1},...,X_{N}}_{2Body}(\rho)\le\Delta_{C}+C\sum_{i\neq j}^{N}\sum_{k,l=1}^{M_{i},M_{j}}\min\{|X_{i}|,|X_{j}|\}|\beta_{kl}^{(ij)}|e^{-\lambda r_{ij}}.$$
	\end{theorem}

    So, if all parts are far enough, then $\mathcal{B}^{X_{1},...,X_{N}}_{2Body}(\rho)$ will also have an upper bound as close as we want to the local bound. Consequently, we will not be able to see substantial violations of any Bell inequality that only involves correlations of one and two bodies in this kind of states.
 
    Analogously, using \textbf{Lemma \ref{lemma2}} together with the Clustering Theorem for thermal states, that is \textbf{Theorem \ref{dec thermal states}}, we get the following result.
	\begin{theorem}\label{twobody bell inequality thermal states}
		If $\rho(\beta)$ is a thermal state  acting on the lattice with inverse temperature $\beta$ less than a fixed $\beta^*$, then there exist $C,\lambda>0$ such that for every $X_{1},\cdots,X_{N}\subset\Omega$ disjoint regions we have $$\mathcal{B}^{X_{1},...,X_{N}}_{2Body}(\rho(\beta))\le\Delta_{C}+C\sum_{i\neq j}^{N}\sum_{k,l=1}^{M_{i},M_{j}}\min\{|X_{i}|,|X_{j}|\}|\beta_{kl}^{(ij)}|e^{-\lambda r_{ij}}.$$
	\end{theorem}
    Again, if all parts are far apart from each other, we have the same conclusion for at most small violations.
	
	As a final result for two-body Bell's inequalities we have the generalization of \textbf{Theorem \ref{CHSH for product states}}. This generalization is also straightfowrward.
	\begin{theorem}\label{twobody for produc states}
		Suppose that the initial state of the system is a product state, i.e, $\rho(0)=\otimes_{x\in\Omega}\rho_{x}$. Then, there is $C,\lambda, v>0$ such that for every $X_{1},\cdots,X_{N}\subset\Omega$ disjoint regions we have
		\begin{align*}
		\mathcal{B}^{X_{1},...,X_{N}}_{2Body}(\rho(t))&\le  \Delta_{C}\\
		&+C(e^{\lambda vt}-1)\sum_{i\neq j}^{N}\sum_{k,l=1}^{M_{i},M_{j}}|X_{i}||X_{j}||\beta_{kl}^{(ij)}|e^{-\lambda r_{ij}}.
		\end{align*}
	\end{theorem}	

    As we discussed, a general Bell inequality in this scenario can involve a sum of correlators of many bodies. Actually, we can write it arbitrarily by:
    \begin{equation}
        \sum_{n=1}^{N}\sum_{i_{1}\neq\cdots\neq i_{n}=1}^{N}\sum_{k_{1},\cdots,k_{n}=0}^{M_{i_{1}},\cdots,M_{i_{n}}}\gamma_{k_{1},\cdots,k_{n}}^{(i_{1},\cdots,i_{n})}\left<E_{k_1}^{(i_1)}\cdots E_{k_n}^{(i_n)}\right>\le\Delta_{C}.\nonumber
    \end{equation}
 
    Thus, analogous to the previous constructions let us define:
   	\begin{align}\label{general Bell bound}
               &\mathcal{B}_{Bell}^{X_{1},...,X_{N}}(\rho,\textbf{E}^{(1)},\cdots,\textbf{E}^{(N)})\nonumber\\ &=\sum_{n=1}^{N}\sum_{i_{1}\neq\cdots\neq i_{n}=1}^{N}\sum_{k_{1},\cdots,k_{n}=0}^{M_{i_{1}},\cdots,M_{i_{n}}}\gamma_{k_{1},\cdots,k_{n}}^{(i_{1},\cdots,i_{n})}\left<E_{k_1}^{(i_1)}\cdots E_{k_n}^{(i_n)}\right>_{\rho};
   \end{align}
    \begin{equation}
          \mathcal{B}^{X_{1},...,X_{N}}_{Bell}(\rho)=\sup_{\textbf{E}^{(1)},\cdots,\textbf{E}^{(N)}}\mathcal{B}^{X_{1},...,X_{N}}_{Bell}(\rho,\textbf{E}^{(1)},\cdots,\textbf{E}^{(N)}).
    \end{equation}
    Generalization of the previous theorems can be obtained, but first we need to extend the notion of \textit{exponential clustering of correlations} to when we are considering the correlations of many bodies at the same time. The next lemma shows us that the assumptions in \textbf{Definition \ref{state with ecc}} are enough to extend the notion of \textit{exponential clustering of correlations} to the case of correlations between many parts.
    \begin{lemma}\label{ecc general}
    	If $\rho$ is a quantum state acting in $\mathcal{H}_{\Omega}$ with \textit{exponential clustering of correlations} then for any set of disjoint regions $X_{1},\cdots,X_{n}\subset\Omega$ and any set of operators $E_{1},\cdots,E_{n}$ supported at $X_{1},\cdots,X_{n}$ respectively we have 
    	\begin{align*}\label{eq state with ecc}
    	&\left|\left<E_{1}\cdots E_{n}\right>_{\rho}-\left<E_{1}\right>_{\rho}\cdots\left<E_{n}\right>_{\rho}\right|\\
    	&\le \|E_{1}\|\cdots \|E_{n}\|(n-1)|X| Ce^{-\lambda r}.
    	\end{align*}
    	where $C,\lambda>0$ are the same constants from the definition of \textit{exponential clustering of correlations}, $|X|=\max\{|X_{1}|,\cdots,|X_{n}|\}$ and $r=\min r_{ij}$, with $r_{ij}$ being the distance between the regions $X_{i}$ and $X_{j}$.
    \end{lemma}
    With this lemma and the same ideas as before, we can generalize \textbf{Theorem \ref{twobody bell inequality groudstate}} and \textbf{Theorem \ref{twobody bell inequality thermal states}}. Before that, in order not to overcharge the notation we will denote by $\Gamma$ the following sum of constants:
    \begin{equation}
    \Gamma=\sum_{n=1}^{N}\sum_{i_{1}\neq\cdots\neq i_{n}=1}^{N}\sum_{k_{1},\cdots,k_{n}=0}^{M_{i_{1}},\cdots,M_{i_{n}}}(n-1)\left|\gamma_{k_{1},\cdots,k_{n}}^{(i_{1},\cdots,i_{n})}\right|.
    \end{equation}
   	\begin{theorem}\label{general bell inequality groudstate}
   	If $\rho$ is the ground state of a gapped Hamiltonian of the lattice, then there exist $C,\lambda>0$ such that for every $X_{1},\cdots,X_{N}\subset\Omega$ disjoint regions we have 
   	\begin{align*}
   	\mathcal{B}^{X_{1},...,X_{N}}_{Bell}(\rho)\le\Delta_{C}+C|X|\Gamma e^{-\lambda r}.
   	\end{align*}
    \end{theorem}
   
   	\begin{theorem}\label{general bell inequality thermal states}
   	If $\rho(\beta)$ is a thermal state  acting on the lattice with inverse temperature $\beta$ less than a fixed $\beta^*$, then there exist $C,\lambda>0$ such that for every $X_{1},\cdots,X_{N}\subset\Omega$ disjoint regions we have
   	$$\mathcal{B}^{X_{1},...,X_{N}}_{Bell}(\rho(\beta))\le\Delta_{C}+C|X|\Gamma  e^{-\lambda r}.$$
    \end{theorem}
    Thus, if the experiment is carried out in such a way that all the parts are away from each other, no Bell inequality will be significantly violated for these two families of states.
\subsection{Summary of results}

    Summing up, this section contains our results divided  into three categories. 
    
    First, we explored non-locality for spin lattices based on the CHSH inequality. We have seen in \textbf{Theorem \ref{CHSH for groundstate}} that if Alice and Bob's actions are restricted to distant regions on the lattice, then the ground state of a gapped Hamiltonian is unable to significantly violate CHSH. Additionally, in \textbf{Theorem \ref{CHSH for thermal states}} we saw that thermal states have an even more restricted behavior, in fact fixed the measurements of one part, there is a minimum distance between them so that from which it is not possible to see violation of CHSH. Now, In \textbf{Theorem \ref{CHSH for product states}}, we saw how non-local correlations are created in time when the initial system is a product state.

    The second bit is a generalization of the first three theorems to a scenario with more parties and more measurements for each part. In this case, we  restricted our analysis to Bell inequalities that only involve correlators of one and two bodies. The conclusions are the same as those obtained for the previous cases, the exception being the \textbf{Theorem \ref{twobody bell inequality thermal states}} where it is no longer possible to conclude non-violation.
    
    Finally, we dealt with general Bell inequalities. For this, we adopted some simplifications, one of which was to look only at the minimum distance between observables. With that, we enunciate the generalizations of \textbf{Theorems \ref{twobody bell inequality groudstate}} and \textbf{\ref{twobody bell inequality thermal states}}.

    We invite the reader to check out Section \ref{Sec.Preliminaries} and Section \ref{Sec.Proofs}. They contain all the mathematical details and  in-depth proofs for the results we approached above.

\section{Preliminaries}\label{Sec.Preliminaries}
\subsection{Bell Inequalities}
    Broadly speaking, in our work we are investigating non-local aspects of spin lattices via Bell-inequalities. Centering our attention on the latter, in this section we cover the basics of what we mean by a non-local correlation.   
    
\begin{figure}
\includegraphics[scale=0.3]{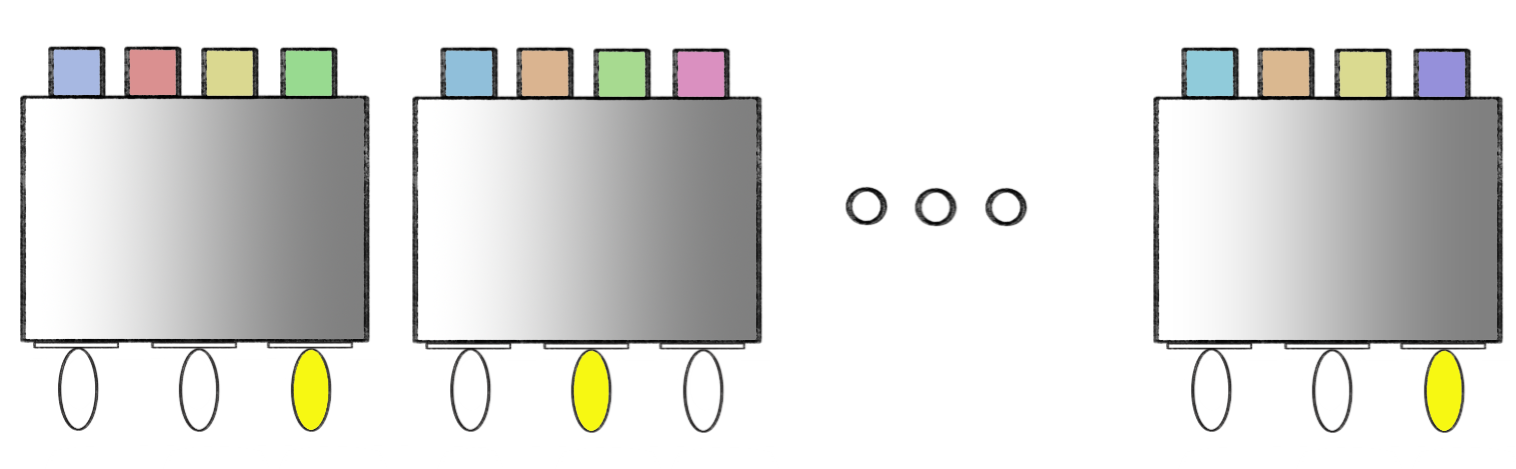}
\caption{Representation for the $(N,m,o)-$scenario. \label{Fig.DeviceIndApproach}}
\end{figure}   
    
    Correlation scenarios are usually formulated in a device-independent language \cite{Bell-Nonlocality}. Think of it as a collection of $N$ black-boxes. Each box comes with $m$ buttons on the top and $o$ light bulbs at the bottom. Whenever a button is pressed, one light bulb goes off as a response to this action. The entire formalism is coined to hidden the inner physical mechanism of each box. As we do not have access to the physical details producing an outcome given that a certain button was pressed, the only description for this $(N,m,o)$-scenario is via the aggregated joint statistics
    \begin{equation}
    \vec{p}=\{p(ab...c | xy...z)\} \in \mathbb{R}^{(om)^N}.
    \label{Eq.DefBehaviour}
    \end{equation}
Each $p(ab...c | xy...z)$ simply means the joint probability of getting outcome $a$ out of the first box when the $x$ button was pressed, and outcome $b$ out of the second box when the $y$ button was pressed, ..., and outcome $c$ out of the $N$th-box when button $z$ was pressed. See Fig.~\ref{Fig.DeviceIndApproach}.
    
The definition of the local set of correlations is motivated in various ways. Particularly, we refer the reader to the recent~\cite{CL14}. For sake of simplicity, we will go with an alternative one. If we assume that each of theses boxes is independent of one another, Eq.~\eqref{Eq.DefBehaviour} would reflect it and factorize as:
    \begin{equation}
    p(ab...c|xy...z)=p(a|x) \times p(b|y) \times ... \times p(c|z)
    \label{Eq.DefIndependence}
    \end{equation}
When correlations across the boxes are detected, and Eq.~\eqref{Eq.DefIndependence} does not hold true, intuitively we assume that what is happening is that there is an exogenous variable, say $\lambda$, we are not accounting for, but that in its presence the independence would manifest:
    \begin{equation}
    p(ab...c|xy...z,\lambda)=p(a|x,\lambda) \times p(b|y,\lambda) \times ... \times p(c|z,\lambda).
    \label{Eq.DefIndependenceLambda}
    \end{equation}
When this is the case, the behaviour we want to look at is nothing but the average of Eq.~\eqref{Eq.DefIndependenceLambda}, in fact:
    \begin{align}
   &p(ab...c|xy...z)=\int_{\Lambda} p(ab...c|xy...z,\lambda) d\mu(\lambda) \nonumber \\
   &=\int_{\Lambda}p(a|x,\lambda) \times p(b|y,\lambda) \times ... \times p(c|z,\lambda)d\mu(\lambda).
    \label{Eq.LocalIntegral}
    \end{align}
Eq.~\eqref{Eq.LocalIntegral} above is the very mathematical expression of what we mean by a correlation to be local. For the finite case it suffices to consider discrete variables, and by doing so we replace the integral for a sum:
    \begin{align}
   &p(ab...c|xy...z)=\sum_{\lambda} p(ab...c|xy...z,\lambda) q(\lambda) \nonumber \\
   &=\sum_{\lambda}p(a|x,\lambda)  p(b|y,\lambda)  ...  p(c|z,\lambda)q(\lambda).
    \label{Eq.DefLocal}
    \end{align}
  In a given $(N,m,o)-$scenario, we say that $p(ab...c|xy...z)$ is \emph{local} whenever it verifies equation ~\eqref{Eq.DefLocal}  above. Basically, it says that there is a hidden-variable we do not have access to that explains the correlation across the boxes. 
    
  For a fixed $(N,m,o)-$scenario, the set of local correlations is a polytope, and as such it can be described through its facets~\cite{Schrijver_2004}. That is to say that every local correlation must satisfy a finite set of linear inequalities. Whenever one of these inequalities is violated, we know for sure that the correlation we are looking at is not attainable with a local model. Because of his influential work on locality, these dividing inequalities are usually known as \emph{Bell-inequalities}~\cite{Bell-Nonlocality}.

    The bipartite correlation scenario, \emph{i.e.} the (2,2,2) scenario, has a single Bell inequality, the CHSH inequality \cite{CHSH}. Satisfying CHSH inequality is a necessary and sufficient condition for a behavior to be local. Note that in more complex scenarios, the local polytope is a multi-faceted object and that in order to attest the a certain correlation is local, we must very all of the facet defining inequalities~\cite{Bell-Nonlocality}.

\subsection{Many-Body Quantum Systems and Clustering Theorems}
    To define our quantum spin system let $\Omega$ be a finite set of sites that will by called by lattice. Let $d$ be a metric in $\Omega$, which gives the distance between the sites in the lattice. We associate to each site $x$ in $\Omega$ a finite-dimensional Hilbert space $\mathcal{H}_{x}$ and for each $X\subset\Omega$ the Hilbert space associate is given by the tensor product $\mathcal{H}_{X}=\otimes_{x\in X}\mathcal{H}_{x}$. The algebra of observables in $X$ is denoted by $\mathcal{L}(\mathcal{H}_{X})$. The support of an operator $A\in\mathcal{L}(\mathcal{H}_{\Omega})$ is given by $\inf\{X\subset\Omega|A=A_{X}\otimes I_{\Omega/X}\}$, where $A_{X}\in\mathcal{L}(\mathcal{H}_{X})$, that is, the support of an operator is given by the smallest set such that the operator acts as an identity in the complement of that set. An interaction for such a system is a map $h$ from the set of subsets of $\Omega$ to $\mathcal{L}(\mathcal{H}_{\Omega})$ such that $h(X)$ has support in $X$. The Hamiltonian is given by $H=\sum_{X\subset\Omega} h(X)$. The dynamics of the model is given by $A(t)=e^{itH}Ae^{-itH}$. Lastly, let $R$ be the maximal distance for the interactions. This is the general construction of a finite quantum spin system. The additional assumption is that the interactions are short-ranged, that is, $R$ is small compared with the size of the lattice.
	
    For this system with specific additional assumptions, there are classes of states that have \textit{exponential clustering of correlations}. The first case is the ground state of a gapped Hamiltonian \cite{communications,Nachtergaele2006-2}. 
	\begin{theorem}[Clustering Theorem for gapped ground state]\label{dec groundstate}
	Let $\rho$ be the ground state of a system with  a spectral gap $\Delta E>0 $ above the ground-energy. Then, exist constants $C,\lambda>0$ such that given two disjoint regions $X,Y\subset \Omega$ at a distance $r$ from each other and two operators $A,B\in\mathcal{L}(\mathcal{H}_{\Omega})$ with support in $X,Y$ respectively, we have the following bound.
    \begin{equation*}
      \left|\left<AB\right>_{\rho}-\left< A\right>_{\rho}\left<B\right>_{\rho}\right| \le \|A\|\|B\||X|Ce^{-\lambda r}.
    \end{equation*} 
	\end{theorem}
    The coefficient $C$ and $\lambda$ are independent of $A$ and $B$. Actually, $C,\lambda$ depends only on the geometry of the lattice, the maximum interaction energy and the spectral gap. For this reason, if we move Alice away from Bob in \textbf{Theorem \ref{CHSH for groundstate}} $\epsilon$ will approach 0. The same argument applies to the \textbf{Theorems \ref{twobody bell inequality groudstate}} and \textbf{\ref{general bell inequality groudstate}}.

    A second class of states with \textit{exponential clustering of correlations} are thermal states. A thermal state, or Gibbs state, of a Hamiltonian $H$ at inverse temperature $\beta$ is given by
    \begin{equation}
    \rho(\beta)=\frac{e^{-\beta H}}{Tr(e^{-\beta H})}.
    \end{equation}

    There exist a universal inverse critical temperature $\beta^{\ast}$, which is, in particular, independent of the system size, below which correlations decay exponentially. This parameter essentially depending on the typical energy of interaction and the spatial dimension of the lattice (see \cite{kliesch2014locality} for more details).  With this assumption, the following theorem was shown in \cite{kliesch2014locality}.
    \begin{theorem}[Clustering Theorem for thermal states]\label{dec thermal states}
	      Let $\rho(\beta)$ be the thermal state at inverse temperature $\beta<\beta^{*}$. There are constants $C(\beta),\lambda(\beta)$ such that if $X,Y$ are two disjoint regions on the lattice at distance $r$ and $A,B$ operator acting in $X,Y$ respectively, we have  
   	      \begin{equation*}
                 \left|\left<AB\right>_{\rho(\beta)}-\left< A\right>_{\rho(\beta)}\left<B\right>_{\rho(\beta)}\right| \le \|A\|\|B\| C(\beta)e^{-\lambda(\beta) r}.
          \end{equation*} 
    \end{theorem}	
    Once again, distancing $X$ from $Y$ does not result in changes to $ C(\beta), \lambda (\beta) $. Because of that and the fact that $rho(\beta)$ is full rank, it was possible to find a minimum distance in \textbf{Theorem \ref{CHSH for thermal states}} so that, from there, it is not possible to violate CHSH. An analogous argument applies to \textbf{Theorem \ref{twobody bell inequality thermal states}} and \textbf{\ref{general bell inequality thermal states}}.

    It is important to emphasize that Lieb-Robinson's bounds are fundamental in the proof of the two previous theorems \cite{LR}. In this seminar paper, Lieb and Robinson prove that there is a bound for the maximal effective velocity for the propagation of information in a quantum spin system with short-range interactions. Another application of Lieb-Robinson bounds is in the propagation of correlations \cite{Nachtergaele2006}. It is easy to see that if we start with a product state, there will be no correlations between the parts of the system. What was shown in \cite{Nachtergaele2006} is that there is a bound to how much correlation can be created in time. This bound grows exponentially with time but decreases exponentially with distance. Indeed, they prove the following result \cite{Nachtergaele2006}.

    \begin{theorem}[Propagation of Correlations]\label{prop_cor}
    	Let $X,Y$ be disjoint regions of $\Omega$ with $r=d(X,Y)$. Let $A,B\in\mathcal{L}(\mathcal{H}_{\Omega})$ have support in $X,Y\subset\Omega$ , respectively, and $\rho(0)=\otimes_{x\in\Omega}\rho_{x}$ be the initial state of the lattice. Then,
    	\begin{align*}
    	&\left|\left<AB\right>_{\rho(t)}-\left<A\right>_{\rho(t)}\left<B\right>_{\rho(t)}\right|\\
    	&\le \|A\|\|B\||X| |Y|C(e^{\lambda vt}-1)e^{-\lambda r}.\nonumber
    	\end{align*}
    \end{theorem}

   \section{Proofs}\label{Sec.Proofs}
     This section contains the proofs of our main theorems. Results that are known in the literature are not discussed here. Our readers might want to check \cite{communications, Nachtergaele2006-2, kliesch2014locality, Nachtergaele2006} to find proofs for theorems \ref{dec groundstate}, \ref{dec thermal states} and \ref{prop_cor}.
	\subsection{Lemma \ref{lemma1}}

    \textbf{Lemma 1.}  \textit{Given two disjoint regions $X,Y\subset\Omega$ and $\rho$ a quantum state with \textit{exponential clustering of correlations}, then $\rho$ is $\epsilon$-local for CHSH with respect this two regions, where $\epsilon=4|X|Ce^{-\lambda r}$.	}

	 The proof of \textbf{Lemma \ref{lemma1}} follows almost directly from \textbf{Definition \ref{state with ecc}}. Indeed, for each pair of measurements $A_{i}$, $B_{j}$ from Alice and Bob respectively, we have
    \begin{align}
	&\left|\left<A_{i}B_{j}\right>-\left<A_{i}\right>\left<B_{j}\right>\right|\le \|A_{i}\| \|B_{j}\||X|Ce^{-\lambda r}.
	\end{align}
    But, as $A_{i}$ and $B_{j}$ has spectrum in $[-1,1]$, we have $\|A_{i}\|$ and  $\|B_{j}\|$ less or equal to 1. So,
	\begin{equation}\label{aibj}
	\left<A_{i}B_{j}\right>\le \left<A_{i}\right>\left<B_{j}\right>+|X|Ce^{-\lambda r}.
	\end{equation}
	Replacing in \eqref{CHSH in lattice}: 
	\begin{align}
	\mathcal{B}_{CHSH}^{X,Y}(\rho,\textbf{A},\textbf{B})&\le \left<A_{0}\right>\left<B_{0}\right>+\left<A_{0}\right>\left<B_{1}\right>\nonumber\\
	&+\left<A_{1}\right>\left<B_{0}\right>-\left<A_{1}\right>\left<B_{1}\right>+4|X|Ce^{-\lambda r}.
	\label{Eq.ProofLemma1B}
	\end{align}
	Let us denote 
	\begin{align}
	\tilde{\mathcal{B}}_{CHSH}^{X,Y}(\rho,\textbf{A},\textbf{B})&=\left<A_{0}\right>\left<B_{0}\right>+\left<A_{0}\right>\left<B_{1}\right> \nonumber \\
	&+\left<A_{1}\right>\left<B_{0}\right>-\left<A_{1}\right>\left<B_{1}\right>.
	\label{Eq.ProofLemma1BTilde}
	\end{align} 
	Therefore,
	\begin{equation}\label{CHSH bound}
	\mathcal{B}_{CHSH}^{X,Y}(\rho,\textbf{A},\textbf{B})\le\tilde{\mathcal{B}}_{CHSH}^{X,Y}(\rho,\textbf{A},\textbf{B})+4|X|Ce^{-\lambda r}.
	\end{equation}
	The term $\tilde{\mathcal{B}}_{CHSH}^{X,Y}(\rho,\textbf{A},\textbf{B})$ defined in Eq.~\eqref{Eq.ProofLemma1BTilde}  above represents an uncorrelated system and then can be simulated by a classical system. For this reason, this quantity must respect the CHSH inequality, so that $\tilde{\mathcal{B}}_{CHSH}^{X,Y}\le2$. Indeed,
	\begin{align}\label{Eq.Max/CHSH}
	\tilde{\mathcal{B}}_{CHSH}^{X,Y}(\rho,\textbf{A},\textbf{B})&=(\left<A_{0}\right>+\left<A_{1}\right>)\left<B_{0}\right>   \nonumber\\
    &+(\left<A_{0}\right>-\left<A_{1}\right>)\left<B_{1}\right>\nonumber\\
	&\le|\left<A_{0}\right>+\left<A_{1}\right>|+|\left<A_{0}\right>-\left<A_{1}\right>|\nonumber\\
	&=2\max\{|\left<A_0\right>|,|\left<A_1\right>|\}\le 2.
	\end{align}
    Now, putting all these elements together in ineq.~\eqref{Eq.ProofLemma1B} we get
	\begin{equation}
	\mathcal{B}_{CHSH}^{X,Y}(\rho,\textbf{A},\textbf{B})\le 2+4|X|Ce^{-\lambda r}.
	\end{equation}
    Finally, note that the term $4|X|Ce^{-\lambda r}$ is independent of $\textbf{A},\textbf{B}$ and because of that we can optimize over all pairs $\textbf{A},\textbf{B}$ without having to care for this term:
    \begin{align}
    \mathcal{B}_{CHSH}^{X,Y}(\rho)&=\sup_{\textbf{A},\textbf{B}}\mathcal{B}_{CHSH}^{X,Y}(\rho,\textbf{A},\textbf{B}). \nonumber \\
    & \leq   2+4|X|Ce^{-\lambda r}.
    \label{Eq.ProofLemma1ConclusionEq}
    \end{align}
 Summing up, Eq.~\eqref{Eq.ProofLemma1ConclusionEq} says that $\rho$ is an $\epsilon$-\textit{local} state for CHSH with respect to $X,Y$ where the appropriate $\epsilon$ is given by $4|X|Ce^{-\lambda r}$.
	
\subsection{Theorem \ref{CHSH for thermal states}}

	\textbf{Theorem 2.}	\textit{Let $\rho(\beta)$ be a thermal state acting on the lattice with a inverse temperature $\beta$ less than a fixed $\beta^{\ast}$, and let $\textbf{A}$ be a set of operators acting on $X\subset\Omega$. There is $r^{\ast}>0$ such that given $Y\subset\Omega$ with $r\ge r^{\ast}$ we have $\mathcal{B}_{CHSH}^{X,Y}(\rho(\beta),\textbf{A},\textbf{B})\le2$ for every set of operators $\textbf{B}$ acting on $Y$.}

    To find a proof for Theorem 2, start recalling that	\textbf{Theorem \ref{dec thermal states}} guarantees us that the thermal states show \textit{exponential clustering of correlations}. That is to say that they satisfy the hypothesis of \textbf{Lemma \ref{lemma1}}. So, from Eq.~\eqref{CHSH bound}:
	\begin{align}\label{CHSHthermal}
	\mathcal{B}_{CHSH}^{X,Y}(\rho(\beta),\textbf{A},\textbf{B})&\le\tilde{\mathcal{B}}_{CHSH}^{X,Y}(\rho(\beta),\textbf{A},\textbf{B})+4|X|Ce^{-\lambda r},
	\end{align}
	where, now, the expected values are obtained with the Gibbs state. 
	
	Once again, we have an upper bound for CHSH as close as we want to the local bound, as long as we consider the parts sufficiently distant from one another. Nonetheless, we can go beyond that and guarantee non-violation. For this, we will use the thermal state property to be full rank. 

	Alice's measurements are dichotomics, so that for each $i\in\{0,1\}$ there is a POVM $\{E_{i}^{(-1)},E_{i}^{(1)}\}$ such that $A_{i}=E_{i}^{(1)}-E_{i}^{(-1)}$. It is known that if Alice's pair of measurements commute, they will not violate CHSH \cite{Bell-Nonlocality}. Therefore, if $A_{i}=\pm \openone$, being $\openone$ the identity operator, then for every quantum state CHSH inequality will not be violated. Hence , suppose  $A_{0}$ and $A_{1}$ different from $\pm \openone$. Thus, neither $E_{i}^{(1)} = 0$ nor $ E_{i}^{(-1)}=0$ holds true. Other than that, as $\rho(\beta)$ is a full rank matrix and also a density matrix, it follows that $\rho(\beta)$ is definite positive. Additionally, we also have that  $E_{i}^{(1)}$ and $E_{i}^{(-1)}$ are positive semi-definite non-null. Then $Tr\left(\rho(\beta)E_{i}^{(1)}\right)$ and $Tr\left(\rho(\beta)E_{i}^{(-1)}\right)$ are strictly positive, and smaller than 1. 
	For this reason,
	\begin{align}
	\left<A_{i}\right>&=Tr\left(\rho(\beta) E_{i}^{(1)}\right)-Tr\left(\rho(\beta) E_{i}^{(-1)}\right)\nonumber\\
	&<Tr\left(\rho(\beta) E_{i}^{(1)}\right)< 1,
	\end{align} 
	and
	\begin{align}
	-\left<A_{i}\right>&=-Tr\left(\rho(\beta) E_{i}^{(1)}\right)+Tr\left(\rho(\beta) E_{i}^{(-1)}\right)\nonumber\\
	&<Tr\left(\rho(\beta) E_{i}^{(-1)}\right)< 1.
	\end{align} 
	Using this fact in \eqref{Eq.Max/CHSH}:
	\begin{align}
	\tilde{\mathcal{B}}_{CHSH}^{X,Y}(\rho(\beta),\textbf{A},\textbf{B})\le 2\max\{|\left<A_0\right>|,|\left<A_1\right>|\}<2,
	\end{align} 
	which shows that there is $\delta>0$ such that: 
	\begin{equation}
	\tilde{\mathcal{B}}_{CHSH}^{X,Y}(\rho(\beta),\textbf{A},\textbf{B})\le 2-\delta.
	\end{equation}
	Therefore, replacing in \eqref{CHSH bound}, we have:
	\begin{equation}
	\mathcal{B}_{CHSH}^{X,Y}(\rho(\beta),\textbf{A},\textbf{B})\le2-\delta+4|X|Ce^{-\lambda r}.\nonumber
	\end{equation}
	Note that so far we have not used anything about Bob's operators. If Bob is far, or to be more exact, if $r\ge r^{*}$, where $r^{*}=\frac{1}{\lambda}\ln\left(\frac{4|X|C}{\delta}\right)$, then $\mathcal{B}_{CHSH}^{X,Y}(\rho(\beta),\textbf{A},\textbf{B})\le2$. Thus, we will not see any violation of CHSH for thermal states as long as we take the measurements far enough. 
\subsection{Theorem \ref{CHSH for product states}}
     
    \textbf{Theorem 3.}	 \textit{ 	Suppose the initial state of the system is a product state, i.e, $\rho(0)=\otimes_{x\in\Omega}\rho_{x}$. Then, there is $C,v,\lambda>0$ such that given two disjoint regions $X,Y\subset\Omega$ then $\rho(t)$ is $\epsilon$-local for CHSH with respect these two regions, where $\epsilon=4|X||Y|C(e^{\lambda vt}-1) e^{-\lambda r}$.}

    The proof of this theorem is completely analogous to that of \textbf{Lemma \ref{lemma1}}. Indeed, for \textbf{Theorem \ref{prop_cor}} we have:
    \begin{equation}
        \left<A_{i}B_{j}\right>\le \left<A_{i}\right>\left<B_{j}\right>+|X||Y|C(e^{\lambda vt}-1) e^{-\lambda r}.
    \end{equation}
	
    Thus, applying the same ideas used from equation \eqref{aibj}, the result is concluded.
\subsection{Lemma \ref{lemma2}}

	\textbf{Lemma 2.}	\textit{Let $\rho$ be a quantum state acting on $\mathcal{H}_{\Omega}$ showing \textit{exponential clustering of correlations}. Then there exist $C,\lambda>0$ such that for every $X_{1},\cdots,X_{N}\in\Omega$ disjoint regions we have $$\mathcal{B}^{X_{1},...,X_{N}}_{2Body}(\rho)\le\Delta_{C}+C\sum_{i\neq j}^{N}\sum_{k,l=1}^{M_{i},M_{j}}|X_{i}||\beta_{kl}^{(ij)}|e^{-\lambda r_{ij}}.$$}

   The proof we discuss in this section follows the same argument we used to prove our first lemma. Using the definition of a state with \textit{exponential clustering of correlations}, for each pair of measurements from different agents, we get: 
   
	\begin{align}\label{ekel}
	\left<E_{k}^{(i)}\right>\left<E_{l}^{(j)}\right>-C|X_{ij}|e^{-\lambda r_{ij}}&\le \left<E_{k}^{(i)}E_{l}^{(j)}\right>\nonumber\\
	&\le  \left<E_{k}^{(i)}\right>\left<E_{l}^{(j)}\right>+C|X_{ij}|e^{-\lambda r_{ij}},
	\end{align}
	for all $i,j\in\{1,\cdots,N\}$ with $i\neq j$, where $|X_{ij}|=\min\{|X_{i}|,|X_{j}|\}$.  Thus, given $\mu\ge0$:
	\begin{align}
	\mu\left<E_{k}^{(i)}E_{l}^{(j)}\right> &\le  \mu\left<E_{k}^{(i)}\right>\left<E_{l}^{(j)}\right>+\mu C|X_{ij}|e^{-\lambda r_{ij}}\nonumber\\
	&=\mu\left<E_{k}^{(i)}\right>\left<E_{l}^{(j)}\right>+|\mu| C|X_{ij}|e^{-\lambda r_{ij}}.
	\end{align}
	On the other hand, if $\mu<0$:
	\begin{align}
	\mu\left<E_{k}^{(i)}E_{l}^{(j)}\right>&\le  \mu\left<E_{k}^{(i)}\right>\left<E_{l}^{(j)}\right>-\mu C|X_{ij}|e^{-\lambda r_{ij}}\nonumber\\
	&=\mu\left<E_{k}^{(i)}\right>\left<E_{l}^{(j)}\right>+|\mu| C|X_{ij}|e^{-\lambda r_{ij}}.
	\end{align}
	So, for every $\mu\in\mathbb{R}$:
	\begin{equation}
	\mu\left<E_{k}^{(i)}E_{l}^{(j)}\right>\le \mu\left<E_{k}^{(i)}\right>\left<E_{l}^{(j)}\right>+|\mu| C|X_{ij}|e^{-\lambda r_{ij}}.
	\end{equation}
	Applying this inequality in Eq. \eqref{Bell bound}, we have
	\begin{align*}
	&\mathcal{B}_{2Body}^{X_{1},...,X_{N}}(\rho,\textbf{E}^{(1)},\cdots,\textbf{E}^{(N)})=\sum_{i,k=1}^{N,M_{k}}\alpha_{k}^{(i)}\left<E_{k}^{(i)}\right>\nonumber\\
	&+\sum_{i\neq j}^{N}\sum_{k,l=1}^{M_{i},M_{j}}\beta_{kl}^{(ij)}\left<E_{k}^{(i)}\right>\left<E_{l}^{(j)}\right>\\
	&+C\sum_{i\neq j}^{N}\sum_{k,l=1}^{M_{i},M_{j}}|X_{ij}||\beta_{kl}^{(ij)}|e^{-\lambda r_{ij}}.
	\end{align*}{}
    Define:
	\begin{align*}
	&\tilde{\mathcal{B}}_{2Body}^{X_{1},...,X_{N}}(\rho,\textbf{E}^{(1)},\cdots,\textbf{E}^{(N)})=\sum_{i,k=1}^{N,M_{k}}\alpha_{k}^{(i)}\left<E_{k}^{(i)}\right>\nonumber\\
	&+\sum_{i\neq j}^{N}\sum_{k,l=1}^{M_{i},M_{j}}\beta_{kl}^{(ij)}\left<E_{k}^{(i)}\right>\left<E_{l}^{(j)}\right>.
	\end{align*}
	So again, $\tilde{\mathcal{B}}_{2Body}^{X_{1},...,X_{N}}(\rho,\textbf{E}^{(1)},...,\textbf{E}^{N})$ represents an uncorrelated system and as such, it must respect the local bound. Therefore,
	\begin{align}
    &\mathcal{B}_{2Body}^{X_{1},...,X_{N}}(\rho,\textbf{E}^{(1)},...,\textbf{E}^{N})\le \Delta_{C}   \nonumber\\
    &+C\sum_{i\neq j}^{N}\sum_{k,l=1}^{M_{i},M_{j}}|X_{ij}||\beta_{kl}^{(ij)}|e^{-\lambda r_{ij}}.
	\end{align}
	Note the right side of the inequality does not depend on what measurements have been taken. Therefore, taking the supreme over them, we have:
	\begin{align}
	&\mathcal{B}_{2Body}^{X_{1},...,X_{N}}(\rho)\le \Delta_{C}+C\sum_{i\neq j}^{N}\sum_{k,l=1}^{M_{i},M_{j}}|X_{ij}||\beta_{kl}^{(ij)}|e^{-\lambda r_{ij}}.
	\end{align}
\subsection{Theorem \ref{twobody for produc states}}
   
   \textbf{Theorem 6.} \textit{Suppose that the initial state of the system is a product state, i.e, $\rho(0)=\otimes_{x\in\Omega}\rho_{x}$. Then, there is $C,\lambda, v>0$ such that for every $X_{1},\cdots,X_{N}\subset\Omega$ disjoint regions we have $$\mathcal{B}^{X_{1},...,X_{N}}_{2Body}(\rho(t))\le  \Delta_{C}+C\sum_{i\neq j}^{N}\sum_{k,l=1}^{M_{i},M_{j}}|X_{i}||X_{j}||\beta_{kl}^{(ij)}|e^{\lambda(vt-r_{ij})}.$$}

    As with \textbf{Theorem \ref{CHSH for product states}}, the proof of this theorem is analogous to that of \textbf{Lemma \ref{lemma2}}. Indeed, for \textbf{Theorem \ref{prop_cor}} we have:
\begin{equation}
\left<E_{k}^{(i)}E_{l}^{(j)}\right>\le  \left<E_{k}^{(i)}\right>\left<E_{l}^{(j)}\right>+|X_{i}||X_{j}|C(e^{\lambda vt}-1)e^{-\lambda r_{ij}}.
\end{equation}

Thus, applying the same ideas used from equation \eqref{ekel}, the result is concluded.
\subsection{Lemma \ref{ecc general}}	
   
    \textbf{Lemma 3.}  \textit{If $\rho$ is a quantum state acting in $\mathcal{H}_{\Omega}$ with \textit{exponential clustering of correlations}, then for any set of disjoint regnions $X_{1},\cdots,X_{n}\subset\Omega$ and any set of operators $E_{1},\cdots,E_{n}$ supported at $X_{1},\cdots,X_{n}$ respectively we have 
    	\begin{align*}\label{eq state with ecc}
    	&\left|\left<E_{1}\cdots E_{n}\right>_{\rho}-\left<E_{1}\right>_{\rho}\cdots\left<E_{n}\right>_{\rho}\right|\\
    	&\le \|E_{1}\|\cdots \|E_{n}\|(n-1)|X| Ce^{-\lambda r}.
    	\end{align*}
    	where $C,\lambda>0$ are the same constants from the definition of \textit{exponential clustering of correlations}, $|X|=\max\{|X_{1}|,\cdots,|X_{n}|\}$ and $r=\min r_{ij}$, with $r_{ij}$ being the distance between the regions $X_{i}$ and $X_{j}$.}  
   
        As $\rho$ is a state with exponential clustering of correlation, we have that:
    \begin{equation}\label{segDEC}
      \left|\left<E_{i}E_{j}\right>-\left<E_{i}\right>\left<E_{j}\right>\right|\le C|X_{i}|e^{-\lambda r_{ij}} \le C|X|e^{-\lambda r},
    \end{equation} 
    for all $i,j\in\{1,\cdots,n\}$. But, more than that, as the minimum distance between the supports of the observables is $r$, given $\{ E_{i_{1}} , \cdots , E_{i_{k}} \} $, we have that $ E_{i_{1}} \cdots E_{i_{k-1}} $ is an observable supported in $ X_{1} \bigcup \cdots \bigcup X_{k-1} $ and the distance between this larger region and the $ X_{{k}} $ is still at least $ r $. Thus, again by the definition of a state with \textit{exponential clustering of correlations}:
    \begin{equation}\label{segDEC2}
      \left|\left<E_{i_1}\cdots E_{i_{k}}\right>-\left<E_{i_1}\cdots E_{i_{k-1}}\right>\left<E_{i_{k}}\right>\right| \le C|X|e^{-\lambda r}.
    \end{equation} 
From this result, we will show by induction that for every $ k \in \{1,\cdots,n\} $:
    \begin{equation}
    \left|\left<E_{i_1}\cdots E_{i_{k}}\right>-\left<E_{i_1}\right>\cdots\left< E_{i_{k}}\right>\right| \le (k-1)C|X|e^{-\lambda r}.
    \end{equation} 
    The case $ k = 1 $ is trivial and the case $ k = 2 $ follows from \eqref{segDEC}. Suppose then, by induction, that the result is true for $ k = m-1 <n $, that is:
    \begin{align}
    \left|\left<E_{i_1}\cdots E_{i_{m-1}}\right>-\left<E_{i_1}\right>\cdots\left< E_{i_{m-1}}\right>\right|\le (m-2)C|X|e^{-\lambda r}.
    \end{align}
    Multiplying this equation by $\left<E_{i_m} \right>$, we have:
    \begin{align}
    &-(m-2)\left|\left<E_{i_m}\right>\right|C|X|e^{-\lambda r} \nonumber \\ 
    &\le \left<E_{i_1}\cdots E_{i_{m-1}}\right>\left<E_{i_{m}}\right>-\left<E_{i_1}\right>\cdots\left< E_{i_{m}}\right> \nonumber\\
    &\le (m-2)\left|\left<E_{i_m}\right>\right|C|X|e^{-\lambda r}.
    \end{align}
    Recalling that $-1\le\left<E_{i_m}\right> \le 1$, we can get rid of each $\left<E_{i_m}\right>$ in the chain above:
    \begin{align}\label{ind1}
      &-(m-2)C|X|e^{-\lambda r}\nonumber\\
      &\le \left<E_{i_1}\cdots E_{i_{m-1}}\right>\left<E_{i_{m}}\right>-\left<E_{i_1}\right>\cdots\left< E_{i_{m}}\right>\nonumber\\
      &\le (m-2)C|X|e^{-\lambda r}.
    \end{align}
    From \eqref{segDEC2} we have
    \begin{align}\label{ind2}
      -C|X|e^{-\lambda r}&\le \left<E_{i_1}\cdots E_{i_{m}}\right>-\left<E_{i_1}\cdots E_{i_{m-1}}\right>\left<E_{i_{m}}\right>\nonumber\\
      & \le C|X|e^{-\lambda r}.
    \end{align}
   Thus, adding \eqref{ind1} and \eqref{ind2}, we have
    \begin{align}
    -(m-1)C|X|e^{-\lambda r} &\le \left<E_{i_1}\cdots E_{i_{m}}\right>-\left<E_{i_1}\right>\cdots\left< E_{i_{m}}\right>\nonumber\\
    &\le (m-1)C|X|e^{-\lambda r}.
    \end{align}
    That is,
    \begin{equation}
      \left|\left<E_{i_1}\cdots E_{i_{m}}\right>-\left<E_{i_1}\right>\cdots\left< E_{i_{m}}\right>\right|\le (m-1)C|X|e^{-\lambda r}.
    \end{equation}
   And so, we concluded the result by induction.

\subsection{Theorem \ref{general bell inequality groudstate}}
    
    \textbf{Theorem 7.}  \textit{If $\rho$ is the ground state of a gapped Hamiltonian of the lattice, then there exist $C,\lambda>0$ such that for every $X_{1},\cdots,X_{N}\subset\Omega$ disjoint regions we have 
    $$\mathcal{B}^{X_{1},...,X_{N}}_{Bell}(\rho)\le\Delta_{C}+C|X|\Gamma e^{-\lambda r}.$$}
    
    From \textbf{Theorem \ref{dec groundstate}} and \textbf{Lemma \ref{ecc general}} we have:
    \begin{align*}
    \gamma_{k_{1},\cdots,k_{n}}^{(i_{1},\cdots,i_{n})}\left<E_{k_1}^{(i_1)}\cdots E_{k_n}^{(i_n)}\right>&\le \gamma_{k_{1},\cdots,k_{n}}^{(i_{1},\cdots,i_{n})}\left<E_{k_1}^{(i_1)}\right>\cdots \left<E_{k_n}^{(i_n)}\right>\\
    &+|X|Ce^{-\lambda r}(n-1)\left|\gamma_{k_{1},\cdots,k_{n}}^{(i_{1},\cdots,i_{n})}\right|.
    \end{align*}
    Then, applying this inequality in \eqref{general Bell bound}, we have
    \begin{align}
    &\mathcal{B}_{Bell}^{X_{1},...,X_{N}}(\rho,\textbf{E}^{(1)},\cdots,\textbf{E}^{(N)})\le    \nonumber\\
    &\sum_{n=1}^{N}\sum_{i_{1}\neq\cdots\neq i_{n}=1}^{N}\sum_{k_{1},\cdots,k_{n}=0}^{M_{i_{1}},\cdots,M_{i_{n}}}\gamma_{k_{1},\cdots,k_{n}}^{(i_{1},\cdots,i_{n})}\left<E_{k_1}^{(i_1)}\right>\cdots\left< E_{k_n}^{(i_n)}\right>   \nonumber\\
    &+C|X|e^{-\lambda r}\sum_{n=1}^{N}\sum_{i_{1}\neq\cdots\neq i_{n}=1}^{N}\sum_{k_{1},\cdots,k_{n}=0}^{M_{i_{1}},\cdots,M_{i_{n}}}(n-1)\left|\gamma_{k_{1},\cdots,k_{n}}^{(i_{1},\cdots,i_{n})}\right|.
    \end{align}
    Let us define:
    \begin{align}
    &\tilde{\mathcal{B}}_{Bell}^{X_{1},...,X_{N}}(\rho,\textbf{E}^{(1)},\cdots,\textbf{E}^{(N)})=   \nonumber\\
    &\sum_{n=1}^{N}\sum_{i_{1}\neq\cdots\neq i_{n}=1}^{N}\sum_{k_{1},\cdots,k_{n}=0}^{M_{i_{1}},\cdots,M_{i_{n}}}\gamma_{k_{1},\cdots,k_{n}}^{(i_{1},\cdots,i_{n})}\left<E_{k_1}^{(i_1)}\right>\cdots\left< E_{k_n}^{(i_n)}\right>.
    \end{align}
    So again, $\tilde{\mathcal{B}}_{Bell}^{X_{1},...,X_{N}}(\rho,\textbf{E}^{(1)},...,\textbf{E}^{N})$ represents an uncorrelated system and as such the local bound must be preserved. Therefore,
    \begin{align}\label{50}
    &\mathcal{B}_{Bell}^{X_{1},...,X_{N}}(\rho,\textbf{E}^{(1)},...,\textbf{E}^{N})\le \Delta_{C}+C|X|\Gamma e^{-\lambda r},
    \end{align}
    where
    \begin{equation}
    \Gamma=\sum_{n=1}^{N}\sum_{i_{1}\neq\cdots\neq i_{n}=1}^{N}\sum_{k_{1},\cdots,k_{n}=0}^{M_{i_{1}},\cdots,M_{i_{n}}}(n-1)\left|\gamma_{k_{1},\cdots,k_{n}}^{(i_{1},\cdots,i_{n})}\right|.
    \end{equation}
    The right side of the inequality \eqref{50} does not depend on what measurements have been taken. Therefore, taking the supreme over them, we have.
    \begin{align}
    &\mathcal{B}_{Bell}^{X_{1},...,X_{N}}(\rho)\le \Delta_{C}+C|X|\Gamma e^{-\lambda r}.
    \end{align}

\subsection{Theorem \ref{general bell inequality thermal states}}
    
    \textbf{Theorem 8.}  \textit{If $\rho(\beta)$ is a thermal state  acting on the lattice with inverse temperature $\beta$ less than a fixed $\beta^*$, then there exist $C,\lambda>0$ such that for every $X_{1},\cdots,X_{N}\subset\Omega$ disjoint regions we have
    $$\mathcal{B}^{X_{1},...,X_{N}}_{Bell}(\rho(\beta))\le\Delta_{C}+C|X|\Gamma  e^{-\lambda r}.$$}
    
    The proof of this theorem is the same as that of the previous one, the only change is in the use of \textbf{Theorem \ref{dec thermal states}} in place of \textbf{Theorem \ref{dec groundstate}}.

\section{Conclusion}\label{Sec.Conclusion}
   In this paper we investigated non-local aspects of many-body quantum systems.  More precisely, using clustering theorems, we demonstrated that relevant classes of quantum states are unable to signal non-locality considerably.  
   
First, exploring the CHSH scenario for spin-lattices, we were able to show that for agents acting only on distant regions of the lattice, the ground state of gapped Hamiltonians cannot exhibit significant violations of any Bell-inequality. Second, we also managed to prove that for thermal states this behaviour is even more restrictive, as there is a minimum distance between regions that screens-off any non-local effect. Finally, we discussed how come non-local correlations evolve in time when the initial state is a simple product state. 

Scenarios with more parties and more measurements were also investigated, but to generalize the results above we had to focus only on Bell-inequalities involving one-body and two-body correlators. A more complete generalization is given in thm.~\ref{general bell inequality groudstate} and ~thm.\ref{general bell inequality thermal states}, and we invite the reader to check them out.

It is natural to ask why we have assumed gapped systems to begin with. Remarkably enough, there is an example in the literature showing this is an assumption we needed to demand. In ref.~\cite{GETELINA20182799}, the ground state of a lattice with short-range interactions is considered, and it is observed that pairs of distant sites do have a violation of CHSH close to the quantum bound. We hope the mathematical toolbox we have provided here can be used to explain why there is this discrepancy between gapped and un-gapped systems when considering non-local aspects.  

Speaking of further works, considering open quantum systems rather than closed ones as we did here, in ref.~\cite{kastoryano_rapid_2013} the authors generalized both the clustering theorem for gapped ground states (thm.~\ref{dec groundstate} above) and the propagation of correlations (thm.~\ref{prop_cor} above). Because these were basic results we built our results upon, we believe that it is also possible to restate thms.~\ref{CHSH for groundstate}, \ref{CHSH for product states}, \ref{twobody bell inequality groudstate}, \ref{twobody for produc states}, and \ref{general bell inequality groudstate} into the framework of open quantum systems. For being more realistic, functioning also a toy-model for quantum memories, we believe that the local aspects of shown by our results can become even more pronounced in this new scenario.

In conclusion, the main message we wanted to put out with this work is that under certain physical assumptions, a large family of quantum systems with many parts behave as classical, local systems. We hope this paper can make a bridge between rather abstract foundations of quantum physics and more palpable many-body quantum physics. This interchange benefiting both areas. 

\begin{acknowledgments}

Many thanks to R. Rabelo for all the discussions. We acknowledge the support from the Brazilian agencies Conselho Nacional de Desenvolvimento Cient\'{i}fico e Tecnol\'{o}gico, Coordena\c{c}\~ao de Aperfei\c{c}oamento de Pessoal de N\'{i}vel Superior, and FAEPEX. This project/research
was supported by grant number FQXi-RFP-IPW-1905 from the Foundational Questions Institute and Fetzer Franklin Fund, a donor advised fund of Silicon Valley Community Foundation.  CD was supported by a fellowship from the Grand Challenges Initiative at Chapman University. 

\end{acknowledgments}

\bibliography{ref}	
\end{document}